\documentclass[journal=jpcbfk,manuscript=article]{achemso}

\usepackage[utf8]{inputenc} % allow utf-8 input
\usepackage[T1]{fontenc}    % use 8-bit T1 fonts
\usepackage{xcolor}

\graphicspath{ figures }

\title{Molecular signatures of pressure-induced phase transitions in a lipid bilayer}

\author{Yanna Gautier}
\affiliation{Université Paris Cité, CNRS, Laboratoire de Biochimie Théorique UPR 9080, 75005, Paris, France}
\alsoaffiliation{CPCV Lab, Department of Chemistry\\
  École Normale Supérieure - PSL University, Sorbonne University\\
  24 rue Lhomond, 75005 Paris, France}

\author{Guillaume Stirnemann}
\affiliation{CPCV Lab, Department of Chemistry\\
  École Normale Supérieure - PSL University, Sorbonne University\\
  24 rue Lhomond, 75005 Paris, France}
\email{guillaume.stirnemann@ens.psl.eu}

\author{Jérôme Hénin}
\affiliation{Université Paris Cité, CNRS, Laboratoire de Biochimie Théorique UPR 9080, 75005, Paris, France}
\email{jerome.henin@cnrs.fr}

\begin{document}

\maketitle

\begin{abstract}
Understanding how lipid bilayers respond to pressure is essential for interpreting the coupling between membrane proteins and their native environments. Here, we use all-atom molecular dynamics to examine the pressure-temperature behavior of model membranes composed of DMPC or $\Delta$9-cis-PC. Within the studied range (288-308 K, 1-2000 bar), DMPC undergoes a liquid--gel transition, while $\Delta$9-cis-PC remains fluid due to unsaturation. The CHARMM36 force field reproduces experimental boundaries with high fidelity: simulated DMPC transitions deviate by only 5--10K and 100--300bar, and $\Delta$9-cis-PC exhibits no transition. Hysteresis is modest but most pronounced when starting from low-temperature gels. We identify area per lipid, bilayer thickness, and acyl-chain gauche fraction as sensitive phase markers; among these, the gauche fraction provides the most robust signature. Simulations indicate an interdigitated gel is the equilibrium structure under finite-size conditions. However, at low temperature and high pressure, interdigitation decreases, consistent with the experimental lamellar gel phase. This long-lived interdigitation critically impacts standard order parameters, specifically area per lipid and membrane thickness. These results underscore the accuracy of modern force fields and highlight how simulations mechanistically complement experimental studies of pressure-regulated membranes.
\end{abstract}

\newcommand{\jh}[1]{{\color{blue}#1}}
\newcommand{\gs}[1]{{\color{teal}#1}}
\newcommand{\yg}[1]{{\color{magenta}#1}}

% keywords can be removed

\section{Introduction}

Most extant life forms are adapted to the pressure and temperature conditions of the Earth's surface.
%, namely, an average temperature of approximately 290~K and a pressure of 1~bar.
The functional structures of biomolecules, the supramolecular assemblies they form, and the chemical reactions they catalyze have evolved to be optimal within this pressure-temperature range, shaped by both the environment and the complexity of the organism. Since phase behaviors such as folding and assembly are critically sensitive to thermodynamic conditions, any significant deviation from these conditions can have dramatic effects on cellular integrity and function.
However, pressure-temperature conditions vary greatly depending on the environmental context. Throughout evolution, especially during the origins of life, the Earth's pressure and temperature conditions were likely different from those we experience today \cite{Winter_origins_2008}. Even now, extreme environments such as the piezosphere, which includes parts of the lithosphere and deep-sea regions, exhibit pressures as high as 1100–1300~bar \cite{FANG_piezosphere_2010}. Some organisms have adapted to these extreme conditions, and understanding the mechanisms behind such adaptations is essential for grasping the evolutionary trajectory of life \cite{Mota_adaptation_2013F, winnikoff_homeocurvature_2024, MEERSMAN_2013}.

Due to its profound impact on biomolecular integrity, high pressure is also employed in food processing as a method to preserve drinks, dairy products, meat, and fish. High-pressure technology is emerging as a promising alternative to traditional thermal treatments, with the goal of minimizing heat damage and reducing the need for chemical additives and preservatives \cite{RASTOGI_2007}. This method inactivates enzymes and destroys bacteria and microorganisms at ambient or even low temperatures.

Unlike water, which remains liquid under increased pressure, membrane lipids can undergo a transition to gel phases at high pressure. Even in the absence of such transitions, high pressure can significantly affect membrane fluidity. This property is essential for molecular interactions \cite{engelman_membranes_2005}, \cite{MacDonald1984}, hydrophobic interactions \cite{GOSH_2001}, the mobility and function of lipids and proteins, and the ability of membranes to adapt their shape and composition to external conditions \cite{lodish_molecular_2003,BRAGANZA_1986,Zein_Winter_2000,PERIASAMY_2009,KULKARNI_2013,EISENBLATTER_2006,KANGUR_2008}. 

A direct consequence of this property is that pressure serves as a valuable tool for investigating the interplay between lipids and membrane proteins, a concept already articulated in the 70s \cite{trudell_1974} and then exploited using a variety of techniques\cite{WINTER_2019}. Recent applications include nuclear magnetic resonance (NMR) spectroscopy studies providing exquisite atomistic details of how high pressure modulates the conformational landscape of membrane proteins embedded in lipid nanodiscs~\cite{pozza_exploration_2022,Fourel2025}. 

In order to understand these effects, it is important to first characterize the impact of pressure of membranes alone, which has been the subject of numerous experimental studies, for pure lipid bilayer \cite{winter_sans_1989, bonev_effect_1997, DRISCOLL_1991, KRISHNAPRASAD_1987, Brooks2011, brooks_pressure_2014} but also mixtures of phospholipids \cite{Brown_2004}. In particular, the phase diagram of pure- and mixed-lipid bilayers is well-documented, with a specific interest in the fluid-to-gel transitions under either low temperature or high pressure \cite{BOTTNER_1994, winter_effect_2009}.

Very few simulation studies have focused on high-pressure effects on lipid bilayers.
%\gs{An early high-pressure MD simulation study investigated the hydrophobic effect among small organic molecules in water\cite{GOSH_2001}, as part of a broad literature on the physics of protein folding and pressure-induced denaturation.}
A pioneering 2011 study \cite{Chen_2011} investigated DPPC and POPC bilayers under pressures up to 300~bar using the united-atom GROMOS54a7 force field. Coarse-grained simulations published in 2012\cite{Lai_pressure_comp_2012} investigated more systematically the phase behavior of a DPPC bilayer, albeit with the limited chemical precision of the early Martini lipid force field.
Later, high-pressure simulations of an all-atom models of POPC and DOPC bilayers\cite{ding_effects_2017} showcased convincing description of pressure effects on the liquid crystal phase, but did not tackle the phase transition region.

Separately, an abundant literature exists on lateral pressure and surface tension in lipid bilayers,\cite{Cantor1997,Marsh2007} yet the connection between these and the effect of isotropic pressure is far from obvious.
The temperature-induced phase transition is comparatively better documented by simulations using modern models \cite{Khakbaz_2018}.

Here we investigate the following points:
(i) the validity of classical force fields across a broad pressure range, particularly given known limitations in their temperature-dependence\cite{LYUBARTSEV20162483};
(ii) the development of robust descriptors of the phase transition;
(iii) the time convergence of structural features under pressure, and the reversibility of simulated pressure-induced phase transitions;
(iv) the presence of interdigitation of the two leaflets. 

This is illustrated here by comparing two membrane compositions. First, 1,2-dimyristoyl-sn-glycero-3-phosphocholine (DMPC, PC(14:0/14:0)) was used to construct membranes with a thin hydrophobic core, closely mimicking the Gram-negative outer membrane where OmpX is naturally found. DMPC is known to exhibit a fluid to gel phase transition at 297~K at 1~bar\cite{ICHIMORI_1998_DMPC_phasediagram}. 
As a control, we simulated 1,2-dimyristoleoyl-sn-glycero-3-phosphocholine ($\Delta$9-cis-PC, PC(14:1/14:1)), a phospholipid with two 14-carbon, mono-unsaturated acyl chains that does not undergo a phase transition under the investigated pressures and temperatures.

We demonstrate that the CHARMM36 force field \cite{Klauda2010} produces results in good agreement with experimental observations.
The bilayer relaxes in response to new pressure and temperature conditions within a few hundreds of nanoseconds, and we discuss a variety of metrics reflecting these changes.
However, some behaviors, such as interdigitation, are subject to metastability and finite-size biases and that must be carefully accounted for in the interpretation of results.

\section{Methods}

\paragraph{Simulated systems}

Systems were built using the CHARMM-GUI Membrane Builder tools \cite{Charmm-gui-membrane_proteins, Charmm-gui, Charmm-gui_membrane_builder, Charmm-gui_input_namd}. Membranes were modelled as periodic bilayer systems. Pure DMPC bilayers were used as model to study the pressure-induced phase transition. $\Delta$9-cis-PC was used as a control lipid because it is very close to DMPC (they have the same number of carbons in the acyl chain) but has an unsaturation at position 9 in each chain which gives a lipid that does not transition from the liquid to the gel phase in the studied pressure and temperature conditions: DMPC has a transition temperature ($T_m$) of 24°C\cite{Black1981}, while the presence of unsaturations lowers the $T_m$ of $\Delta$9-cis-PC to negative temperatures\cite{Coolbear1983}.
We used the following parameters in the CHARMM-GUI Membrane Builder to construct DMPC and $\Delta$9-cis-PC bilayers: a rectangular box with a dimension along the vertical z axis based on a water thickness of 22.5~Å, and of dimension 60~Å along x and y. Using this parameters, we obtained DMPC bilayers with 57 molecules per leaflet and 54 molecules of $\Delta$9-cis-PC per leaflet in the lipid control systems. Finally, we added a concentration of 0.15 M of NaCl to each system.

\paragraph{Molecular dynamics simulations}

All MD simulations were performed using NAMD version 3 alpha 7 \cite{namd} with the CHARMM36 all-atom lipid force field \cite{Klauda2010} and the TIP3P water model. Simulations were carried out under periodic boundary conditions based on rectangular boxes containing a hydrated lipid bilayer (in the xy plane, i.e., normal to the z-axis), either pure or containing an embedded OmpX protein. We used hydrogen mass repartitioning \cite{hydrogen_mass_repartitioning} combined with RATTLE constraints on the length of bonds involving hydrogen, which allows the use of a 4~fs time step. The systems were first energy-minimized and relaxed in the NPT ensemble (i.e. with a constant number of particles N, constant pressure P and constant temperature T), at 1~bar and 303.15~K. The files provided by the CHARMM-GUI Membrane Builder were used for minimisation, consisting of 6 cycles of 90~ns each, with planar and dihedral restraints that are progressively removed over the cycles. The systems were then additionally relaxed with a time step of 4~fs for 10~ns.

During the production run, a range of pressures (1~bar, 250~bar, 500~bar, 750~bar and 2000~bar) and temperatures (288~K, 298~K and 308~K) were used to explore the DMPC phase diagram. A Langevin thermostat (damping of 1~ps$^-1$) and NAMD's Nosé-Hoover Langevin piston method (period of 200~fs, decay time of 100~fs) \cite{Nose_hoover_method, Langevin_dynamics} were used to control the temperature and pressure, respectively. Particle mesh Ewald was set with a grid spacing of 1.0~Å. Non-bonded interactions were cutoff at 12.0~Å and switching functions were applied beyond 10~Å for both electrostatics and van der Waals forces. DMPC bilayers were simulated for 1~µs, with 3 replicas of each system, for a total simulation time of 36 µs. The control lipid was simulated for 9x1~µs for the pure lipid bilayer. Configurations were saved every 100~ps for analysis.

The target pressures and temperatures were applied at the beginning of the production run as instantaneous change. To check the impact of the protocol on the phase transition, we also tested a smooth transition based on linear ramps: 0.1~K/ns for temperature and 10~bar/ns for pressure. This test was performed for the three replicas under three conditions: 1~bar and 288~K, 250~bar and 298~K, and 750~bar and 308~K. For each condition, the ramp was applied to the initial relaxed bilayer configuration, and the simulation ran for 1~µs after the ramp.

For the T,P condition close to the fluid-to-gel transition limit, we probed the phase of the DMPC bilayer using a starting structure that was half gel and half fluid. To do so we used a DMPC bilayer in the gel phase  obtained from a 1~µs-long MD simulation at 2000~bar and 288~K. Then, we restrained the positions of a cylinder of lipid molecules in the center of the bilayer. This cylinder represents half of the DMPC molecules in the bilayer. The pressure and temperature were then put at 1~bar and 308~K, respectively, to liquefy the DMPC molecules that were not restrained. This half-gel, half-fluid bilayer was then relaxed at 1~bar and 293~K for 500~ns, creating three replicas. 

\paragraph{Trajectory analysis}

Trajectories were analyzed using VMD \cite{VMD} version 1.9.4 through Tcl scripts. The following structural properties were calculated: the volume of the simulation box, the area per lipid, the gauche fraction of the lipid chains, the bilayer thickness, the interdigitation of the lipids in the bilayer, the positions of the last carbon of the tails of the lipids (C14), and the positions of the phosphorus atoms. 

The box volume was normalized to the condition with the largest volume: 1~bar, 308~K. The area per lipid was defined as the area of a leaflet (corresponding to the area of the simulation box in the xy plane) divided by the number of lipids it contains. The dihedral angles of the lipid chains were used to calculate the gauche fraction, which corresponds to the number of gauche angles (grouping the \textit{gauche+}, less than 120 degree, and \textit{gauche-}, more than -120 degree, angles \cite{gauche_angles}) divided by the total number of dihedral angles in the system (11 dihedrals for each lipid chain for all lipids). The gauche fraction was used as an indicator of lipid phase transition because it clearly describes the geometry of the lipids.

We investigated the presence of interdigitations in the lipid bilayer by tracking the distribution of z-positions of terminal carbon atoms (C14) in lipid chains. To better characterize these interdigitations, we examined the distance between the planes formed by the C14 atoms of the top and bottom leaflets. The entire simulation trajectory was centered, aligned, and fitted without changing the membrane's orientation. Lipids were marked as belonging to the "upper" or "lower" leaflet according to their respective positions relative to the bilayer midplane. The x, y, and z positions of the top and bottom C14 atoms were retrieved from the simulations. A 100x100 grid was created for each leaflet using the corresponding C14 atoms. Each grid represents the plan formed by the C14 atoms of each leaflet. Then, the distance between each leaflet was calculated by subtracting the bottom grid from the top one. We fit a Gaussian distribution to the distribution of distances, which allows us to conclude the population of interdigitated lipids in the bilayer (Figure \ref{fig:DMPC_interdigitations_dist}). The main positive-centered peak of the distance distribution corresponds to lipids that are not interdigitated, while the distance values below the main peak correspond to interdigitated regions of the bilayer.
We fit a Gaussian distribution to the main peak, masking the negative values.
The fraction of the distribution not accounted for by the Gaussian kernel corresponds to the bilayer's interdigitated area, we use this as a metric for the fraction of interdigitated lipids in the bilayer. This analysis was performed every 0.1~ns over the last 300~ns of simulation, after the fluid-to-gel transition.

Lipid interdigitations and phase transition can also be witnessed by measuring the bilayer thickness. In the same principle as what was done for the C14 atoms the x, y and z coordinates of each phosphorus atom were extracted for each frame of the simulation. For each leaflet, a 100x100 grid was constructed based on the x, y positions of the phosphorus atoms. Vertical positions of the phosphorus atoms were extrapolated by linear interpolation to create a plane corresponding to the surface of the upper and lower leaflets. The instantaneous distance between the upper and lower planes was then calculated and considered as the bilayer thickness. The bilayer thickness was measured every 0.1~ns for the last 300~ns of the simulation. Movies of the bilayer thickness were made showing the thickness calculated every nanosecond over the entire length of the simulation. These movies can be found in the Supplementary data section.

The volume of the simulation box, as well as the area per lipid, and as the gauche fraction were expressed as a function of time with running averages calculated using a convolution method with a window that will be specified when needed. The averaged presented in this paper were calculated on the data of each replica and the standard deviations were calculated between the different replicas.

\section{Results and discussion}

\subsection{Pressure-induced deformation of a water/DMPC bilayer system}

Our model system is an infinite, planar bilayer of DMPC in water. We chose this lipid because it undergoes a liquid–gel phase transition under the considered (P,T) conditions (1-2000~bar, 288-308~K) \cite{ICHIMORI_1998_DMPC_phasediagram}. These conditions not only reflect those commonly found in natural environments on Earth but also match those used in recent NMR studies investigating how this phase transition influences the behavior of an embedded integral membrane protein~\cite{pozza_exploration_2022,Fourel2025}. 

After a short equilibration at 1~bar and 303.15~K, initial configurations were selected and the thermodynamic conditions adjusted to target values ranging from 1 to 2000~bar for pressure, and from 288~K to 308~K for temperature. For each (P,T) condition, three independent replicas were propagated (see Methods). We note that under these initial thermodynamic conditions, the bilayer remains in a liquid-like state initially.

\begin{figure}[htp]
    \centering
    \includegraphics[width=14cm]{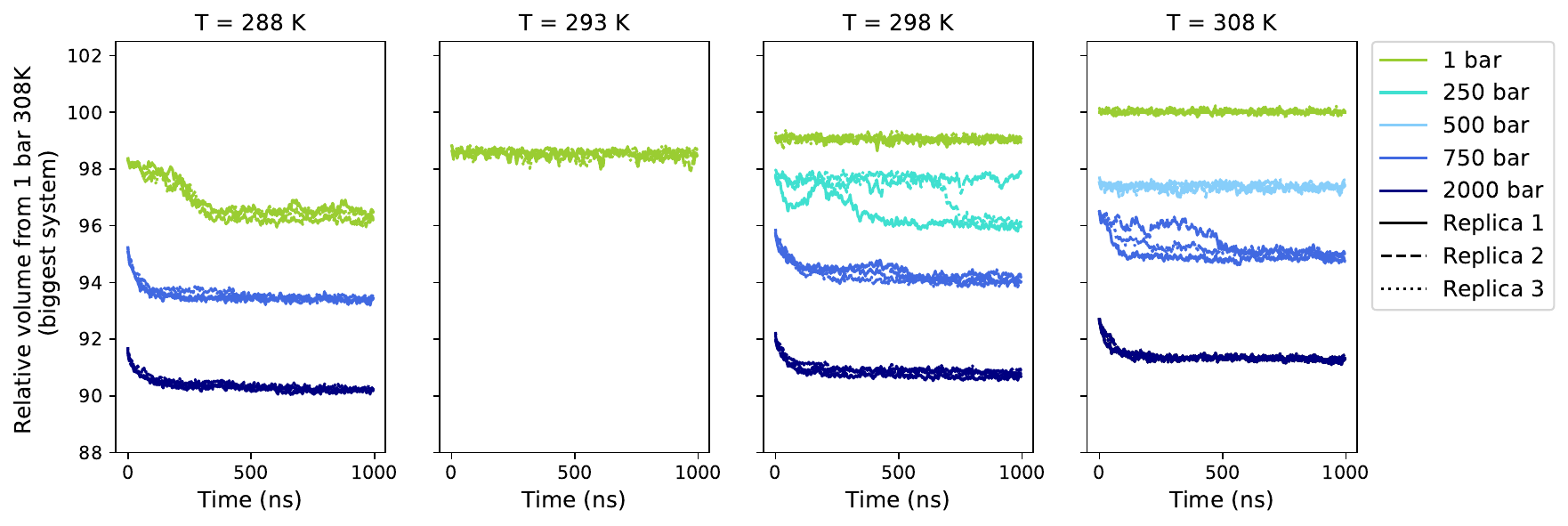}
    \caption{\textbf{Thermodynamics of DMPC bilayer in water}. Volumes have been normalized to the condition with the highest volume among all conditions: 1~bar and 308~K. Values presented here correspond to the running averages computed over 5~ns for all pressures and temperatures conditions simulated.}
    \label{fig:DMPC_box_volume}
\end{figure}

The evolution of the simulation box dimensions reveals that relaxation occurs on two distinct timescales (Figure~\ref{fig:DMPC_box_volume}). At lower temperatures and/or higher pressures, the box volume first decreases rapidly over the course of a few nanoseconds, reflecting equilibrium molar volumes of water and lipids in the fluid phase (Note that this regime is not clearly visible in Figure~\ref{fig:DMPC_box_volume} because data is smoothed by the running averages and the limited time-resolution of the saved trajectory of molecular configurations. However, all curves should in principle start from 100\%). Depending on the thermodynamic conditions, a second, much slower relaxation phase follows, with volume decreasing further over 100--400~ns. While all replicas show consistent behavior under high-pressure, low-temperature conditions, a noticeable variability in relaxation timescales is observed under milder conditions (for example, 308~K and 750~bar, Figure~\ref{fig:DMPC_box_volume}); nonetheless, all replicas eventually converge to the same final volume plateau.

These mesoscopic volume and shape changes are coarse indications of a phase transition from a liquid-disordered to a gel-like state, which we will later characterize at the molecular level. Notably, this behavior is independent of the protocol used to reach specific (P,T) conditions. We verified this by conducting a set of simulations where pressure and temperature were gradually ramped from the initial state at 303.15~K and 1~bar, using fixed rates (Figure~\ref{fig:PT_ramp}), mimicking an experimental protocol—albeit at rates much faster than would be used experimentally, in order to remain tractable within MD simulations. No differences in final behavior were observed between the ramp and instantaneous protocols, suggesting that in this system, phase transitions are relatively insensitive to the specific path taken through (P,T) space. For this reason, we use the instantaneous switching protocol for the remainder of this study.
The overall volume change reflects the compressibility of each phase, as well as the volume difference between phases, which is the thermodynamic driving force behind the pressure-induced transition.
We find volume changes up gelation on the order of -45~\AA$^3$ per lipid at 1~bar (Figure~\ref{fig:DV_perlipid}). This change reduces to -30~\AA$^3$ at 2000~bar, reflecting a higher compressibility of the fluid phase than the gel.
Therefore we can estimate that the work of pressure forces at 250~bar promotes gelation to the amount of 0.1~kcal/mol per lipid.

\subsection{Molecular evidence for the phase transition}

While significant changes in box volume are indicative of a phase transition, the transition itself is more precisely characterized using local descriptor that capture local lipid conformations and structural organization. Among the many possible collective variables, we focus here on three complementary metrics (see Methods). While the concept of \textit{order parameter} is widely used in condensed matter thermodynamics, we deliberately avoid this term here, as it refers to specific variables within the lipid/NMR community.

The first is the area per lipid, which decreases sharply during the transition as the hydrophobic chains align and pack more tightly in the gel phase \cite{ruocco_1982},\cite{blaurock_1986}. The second is the gauche fraction, defined as the proportion of dihedral angles along the acyl chains adopting a gauche conformation. In the liquid-disordered state, lipids display substantial conformational flexibility, whereas the transition to the gel phase involves a marked reduction in gauche conformations in favor of more extended trans conformations. The third parameter is membrane thickness, which increases upon gelation due to the straightening and ordering of the hydrophobic chains.

\begin{figure}[htp]
    \centering
    \includegraphics[width=14cm]{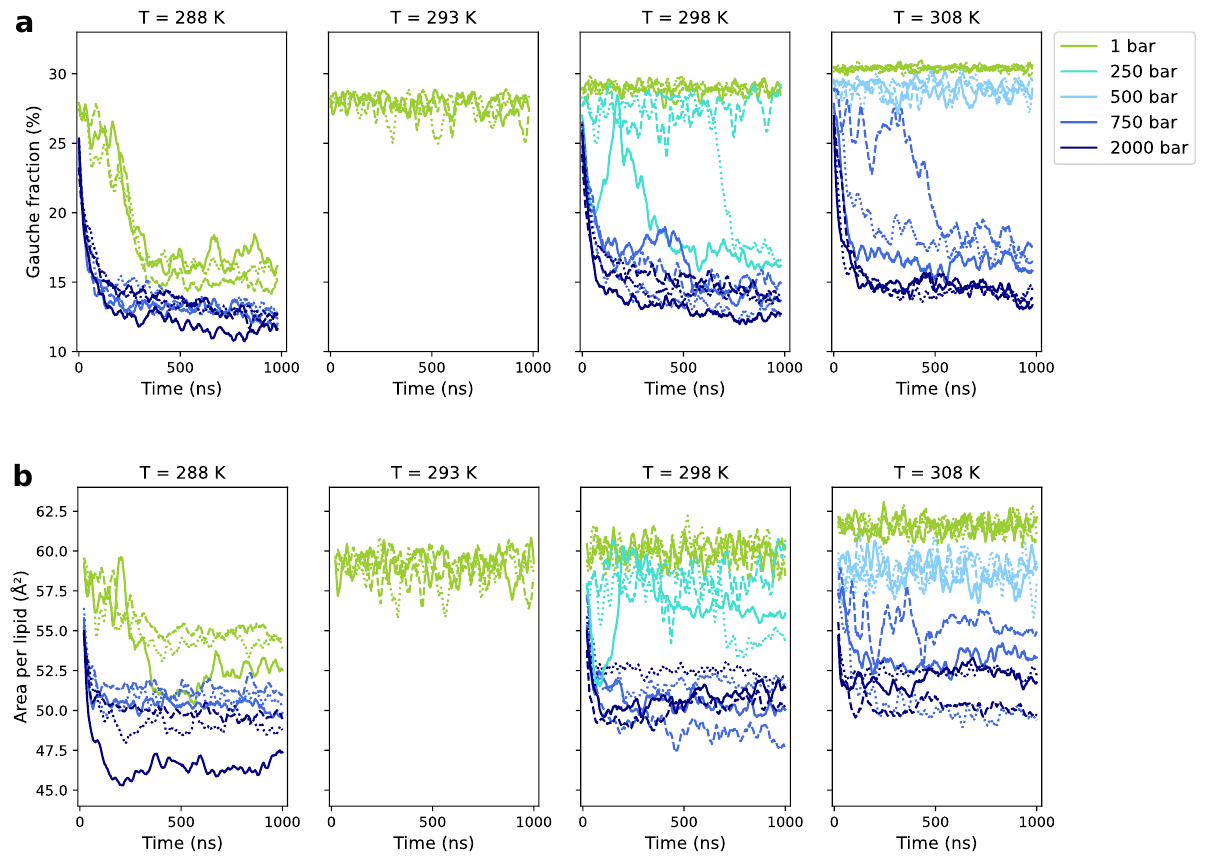}
    \caption{\textbf{Molecular phase transition descriptors.} (a) The gauche fraction of the dihedral angles of the acyl chains of the DMPC lipids and the (b) DMPC area per lipid are shown as a function of time. Each panel corresponds to a different simulated temperature (288~K, 293~K, 298~K and 308~K). The simulated pressures at each temperature are shown on the corresponding panels. The values for each replicas are represented by a different line style with replica 1 in solid, replica 2 in dashed, and replica 3 in dotted style. Values presented here correspond to the running averages computed over 20~ns.} 
    \label{fig:DMPC_area_per_lipid_gauche_fraction}
\end{figure}

Overall, these three descriptors exhibit time evolution similar to that of the system volume (Figure~\ref{fig:DMPC_area_per_lipid_gauche_fraction}). We first examine the gauche fraction (Figure~\ref{fig:DMPC_area_per_lipid_gauche_fraction}a). Under conditions of high temperature and/or low pressure, where the bilayer remains in the liquid phase, little change is observed in the gauche fraction over the microsecond timescale, across all three replicas. In these cases, the gauche fraction increases with temperature and decreases with pressure, reflecting enhanced molecular fluctuations within the liquid crystal phase under such conditions.

In contrast, at low temperatures and/or high pressures, the gauche fraction exhibits a sharp drop within the first 100~ns, followed by a slower relaxation phase. We interpret this rapid decrease as a signature of the liquid-to-gel phase transition, while the slower relaxation reflects further ordering within the gel phase. Notably, under conditions strongly favoring gel formation, all replicas show consistent behavior in both transition timing and final values. However, for intermediate (P,T) conditions, close to the phase boundary (298~K/250~bar, 308~K/750~bar), we observe significant fluctuations in the gauche fraction, including large discrepancies between replicas and oscillations between high and low values over time. This behavior suggests that the system is either dynamically sampling both phases, or undergoing a markedly slower transition, likely due to proximity to a transition line or coexistence region. Importantly, these subtle dynamic features are not evident from volume changes alone (Figure~\ref{fig:DMPC_box_volume}), highlighting the increased sensitivity and informativeness of the gauche fraction as a descriptor for identifying and characterizing lipid phase behavior, in addition to having a physical connection to the NMR observables\cite{gauche,Fourel2025}.

Having shown that the average gauche fraction of dihedral angles along the lipid acyl chains provides a good descriptor of the phase transition from liquid to gel, we next examine how consecutive dihedral angles along the chains exhibit preferences for trans conformations upon gelation (Figure~\ref{fig:Gauche_fraction_by_dihedral}).  The central part of the chain tends to adopt a larger fraction of trans conformations compared to the extremities, i.e., the region in contact with the glycerol head (C1 side, dihedral D1) and the terminal tail (C14 side, up to dihedral D11). Interestingly, this effect is already present in the liquid state and becomes even more pronounced upon gelation (Figure~\ref{fig:Gauche_fraction_by_dihedral}). Notably, the gauche fraction of the last dihedral angle (in contact with the opposite leaflet) is typically twice that of the chain core. On the other side, the perturbation induced by the lipid head propagates along several carbon atoms, spanning almost 1~nm, with the minimum gauche fraction observed around the seventh dihedral (C7 to C10). Therefore, even in the gel phase, parts of the lipid acyl chains remain highly disordered, particularly at the extremities, as expected\cite{petrache_2000, Vermeer_2007}.

Overall, both the area per lipid and membrane thickness exhibit similar trends across the explored (P,T) conditions, reflecting the expected structural transformations associated with the liquid-to-gel phase transition (Figure~\ref{fig:DMPC_area_per_lipid_gauche_fraction}b and Figure~\ref{fig:DMPC_bilayerthickness}). For example, at 288~K, all simulations show a marked decrease in both parameters on the timescale of a few hundred nanoseconds. At 298~K and 250~bar, the system displays clear bistability, with oscillations between liquid and gel phases. However, in some cases we observe significant deviations among replicas (e.g. 288~K and 2000~bar), or an apparent lack of temporal decrease in these parameters despite other indicators of a phase transition (such as 308~K and 750~bar, for which the gauche fraction evolution suggests a final gel-like state, Figure~\ref{fig:DMPC_area_per_lipid_gauche_fraction}a).

\begin{figure}[htp]
    \centering
    \includegraphics[width=14cm]{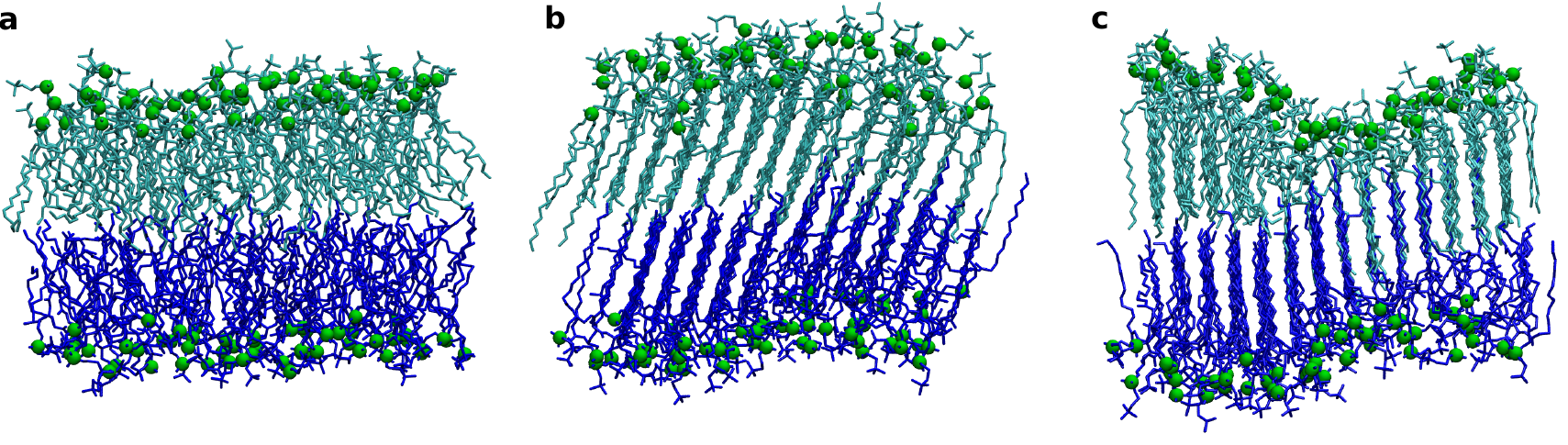}
    \caption{\textbf{Morphology of DMPC bilayers} 
    Molecular rendering of the various DMPC phases and bilayer organizations observed. (a) The crystal-liquid phase, taken here after 1 µs of simulation at 1~bar and 308~K. (b) The gel phase with no major interdigitations, observed after 1 µs of simulation at 750~bar and 288~K. (c) Finally, the interdigitated gel phase, observed in most conditions where the fluid-to-gel transition occurred, here taken after 1 µs of simulation at 2000~bar and 308K. Images rendered with VMD.\cite{VMD}}
    \label{fig:visu}
\end{figure}

Visual inspection of the trajectories (Figure~\ref{fig:visu}) provides insight into the origin of these discrepancies between mesoscopic parameters (area per lipid and membrane thickness), which are collective properties defined at the simulation-box scale, and the gauche fraction, which is a molecular-level metric. Specifically, the gel phases in these simulations often exhibit interdigitation, i.e., interpenetration of the lipid tails from opposing leaflets. For a given average gauche fraction, the degree of interdigitation can substantially influence both membrane thickness and area per lipid. Increased interdigitation introduces vertical fluctuations that may lead to lateral compaction, thereby reducing the area per lipid while complicating straightforward interpretation of membrane thickness.

While these are two metrics which are widely used in the field, they are thus quite sensitive to mesoscopic fluctuations of the bilayer, whereas the gauche fraction appears as a more reliable molecular reporter of the phase transition. In particular, previous studies highlighted that interdigitation was at the origin of the difference in surface area between DPPC mono- and bilayers\cite{Cascales_2006}. In addition, the measure of the averaged area per lipid has a high degree of uncertainty depending of the employed experimental technique\cite{NAGLE_2000}, and the membrane thickness is not straightforwardly measured. 

\subsection{P/T phase diagram of infinite DMPC bilayers and hysteresis}

In order to clearly classify the thermodynamic state of the membrane (liquid, gel, or bistable), we examined the distributions of gauche fractions, which provide clear signatures of the phase transition (Figure~\ref{fig:DMPC_gauche_frac_distrib}). In all but one of the investigated thermodynamic conditions, the distributions are unimodal and centered around two different values which are similar for a given set of replicas, i.e. 10--20\%, or around 30\%. At 298~K and 250~bar, there is significant dispersion among replicas, and one is bimodal, reflecting the variations seen in Figure~\ref{fig:DMPC_area_per_lipid_gauche_fraction}. Based on these observations, we defined a threshold line of 22\% separating the gel from the liquid regions.

\begin{figure}[htp]
    \centering
    \includegraphics[width=18cm]{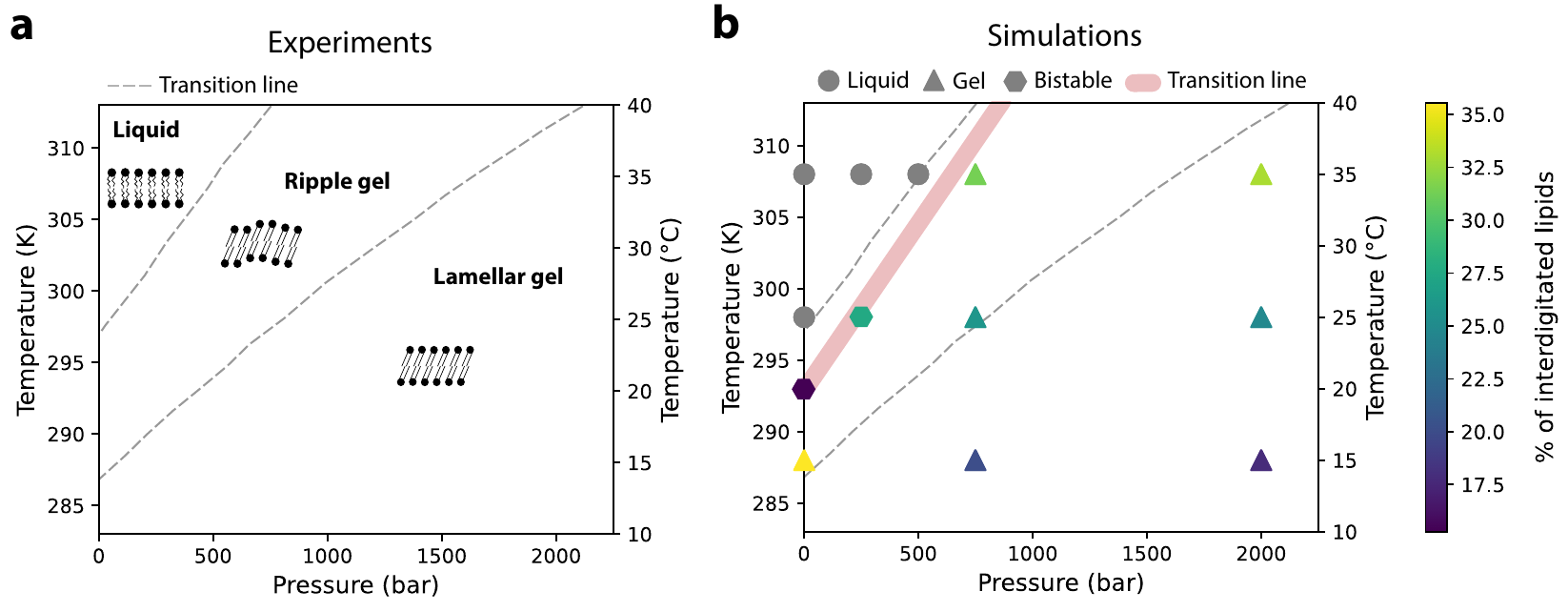}
    \caption{\textbf{DMPC phase diagram}. The experimental phase diagram of DMPC bilayer \textbf{(a)} adapted from Ichimori \textit{et al.}\cite{ICHIMORI_1998_DMPC_phasediagram} shows the experimental melting line between gel and fluid (upper most dashed line) and ripple and lamellar gel (lower dashed line). The DMPC phase diagram built from the MD simulations \textbf{(b)} shows the temperature and pressure conditions under which we conducted our molecular dynamics simulations, starting from a liquid phase. After 1 µs of simulation, we obtained either the crystal liquid phase (indicated by circles), or the gel phase (indicated by triangles), or metastable states where some replicas stayed in the crystal liquid phase and some replicas experienced fluid-to-gel phase transition (indicated by hexagons). Points are colored according to the \% of interdigitated lipids in the gel phase. The values were calculated from the last 300~ns of simulation. The interdigitation values shown here were calculated as the mean over the replicas. The thick pink line corresponds to the fluid-to-gel transition we can build from the simulations. To facilitate comparison with experimental literature, temperature values are presented in both Celsius and Kelvin.}
    \label{fig:phase}
\end{figure}

This classification enables the construction of a DMPC bilayer phase diagram and the approximate identification of a transition line separating gel and liquid regions (Figure~\ref{fig:phase}). For comparison, the experimentally determined melting line is also shown (Figure~\ref{fig:phase}a) \cite{ICHIMORI_1998_DMPC_phasediagram}. The simulated line passes through thermodynamic conditions where bistability is observed (Figure~\ref{fig:phase}b).  The agreement between the simulation results and experimental data is overall quite good. Despite the force field being built on a fragment approach calibrated using \emph{ab initio} calculations, it appears to reliably capture the pressure and temperature dependence of the DMPC bilayer phase behavior.

However, we emphasize that the exact position of the transition line cannot be determined with high precision using the current protocol. First, bistable states may continue to evolve toward one of the two phases over longer simulation timescales, which are currently inaccessible on a routine manner. Second, our approach has so far relied exclusively on irreversible transitions from the liquid to the gel phase, starting from a liquid configuration. To probe hysteresis in the system, we also consider the reverse process—namely, starting from a gel phase that is suddenly heated or depressurized (Figure~\ref{fig:hysteresis}).

For example, at 1~bar, reducing the temperature starting from the liquid phase results, after 1 microsecond of simulation replicated three times, in a gel phase only at 288~K (Figure~\ref{fig:DMPC_area_per_lipid_gauche_fraction}a). For all temperatures above 293~K, the system remains liquid with no clear sign of transition. However, we suspected that the significant fluctuations of the gauche fraction observed at 293~K were indicative of a point close to the transition line (Figure~\ref{fig:DMPC_area_per_lipid_gauche_fraction}a). For this condition, we thus test whether an initial gel/liquid 50:50 mixture relaxes toward a liquid or a gel. As shown in Figure~\ref{fig:DMPC_phase_293K}, two replicas became liquid while one became a gel. This metastability confirms that 293~K appears to lay on the transition line. When now heating this gel at increasing temperatures starting at 293~K, with 5~K increments over the course of more than 2.5~µs, there is no evidence of melting, despite noticeable fluctuations and increase in the gauche fraction (Figure~\ref{fig:hysteresis}a), until 308~K, where the system immediately melts (Figure~\ref{fig:hysteresis}b). 

We then seek to compare hysteresis along either temperature or pressure, starting from a point that experiences a phase transition upon pressurization (298~K, 250~bar) (Figure~\ref{fig:hysteresis}c and d). Upon heating at 308~K, the bilayer melts (Figure~\ref{fig:hysteresis}c), which corresponds to the thermodynamic state observed when the system was initially subject to a the same conditions but starting in the liquid phase (Figure~\ref{fig:phase}). Similarly, reducing the pressure to 1~bar while keeping the same temperature of 298~K is accompanied with rapid melting, while further re-pressurization at 250~bar induces a phase transition again. 

While this investigation was performed on limited conditions, this suggests that some hysteresis is observed, especially when starting from a gel at low temperature, while another condition close to the transition line, for which we sampled the reversibility of the phase transition along either temperature or pressure, showed much more limited asymmetry in the transition behavior.

\subsection{Behavior of a control lipid with two \textit{cis} unsaturated chains}

The emergence of a gel phase for DMPC under \textit{mild} and biologically relevant thermodynamic conditions stands in sharp contrast to the phase behavior of $\Delta$9-cis-PC, which has a melting temperature of approximately 270~K at 1~bar. $\Delta$9-cis-PC shares the same heavy-atom composition and overall molecular structure as DMPC, with the sole difference being a \textit{cis} double bond at the 9th carbon along each hydrophobic tail. This unsaturation disrupts the formation of extended all-trans conformations, thereby impeding the ordering necessary for gelation. As a result, the transition to a gel phase for $\Delta$9-cis-PC requires significantly lower temperatures and/or higher pressures. In previous high-pressure NMR experiments on nanodiscs, $\Delta$9-cis-PC was employed as a control lipid that does not exhibit a phase transition within the 288--308~K and 1--2000~bar range~\cite{pozza_exploration_2022}.

To complement and extend these findings, we repeated our simulation protocol using a periodic bilayer composed of $\Delta$9-cis-PC.
In contrast to the DMPC system, the structural parameters remain stable over time across all (P,T) conditions, see Figure \ref{fig:D9_area_per_lipid_gauche_fraction}. The gauche fraction consistently displays values characteristic of a liquid-disordered phase, with minimal dependence on pressure and a slight increase with temperature, reflecting enhanced molecular fluctuations. Similarly, the area per lipid remains high across all simulations and exhibits a mild decrease with applied pressure. 
%\textcolor{red}{[can we get something like an isothermal compressibility out of it and compare to experimental data? same question for DMPC voir par ex https://www.frontiersin.org/journals/physics/articles/10.3389/fphy.2021.636149/full]}.
As expected, no signatures of a phase transition to a gel state, such as a significant drop in gauche fraction or a reduction in area per lipid, are observed in any $\Delta$9-cis-PC simulations, in agreement with available experimental data.

%%%%%%%%%%%%%%%%%%%%%%%%%%%%%%%%%%%%%%%%%%%%%%%%%%%%%%%%%%%%%%%%%%%%%%%%%%%%%%%%%%%%%%%%%%%%%%%%%%

\subsection{Interdigitation in the gel phase}

We finally turn to the phenomenon of interdigitation, i.e., the interpenetration of lipid tails from opposing leaflets (Figure \ref{fig:visu}). While we have already suggested that this process may underlie some of the variability observed in mesoscopic quantities such as area per lipid and membrane thickness, it can be characterized more directly. Specifically, we examine the distributions of the terminal C14 carbon atoms (the last atom on the hydrophobic tails) along the membrane normal (z-axis), perpendicular to the bilayer plane (Figure~\ref{fig:DMPC_interdigitations}).

In the liquid state, these distributions are unimodal for each leaflet, with well-separated peaks and minimal variation across replicas. The typical inter-leaflet distance, defined as the separation between the two peaks, is approximately 0.3~nm. 

In contrast, under all (P,T) conditions where gelation occurs, the distributions reveal a striking feature: in each leaflet, a sub-population of C14 atoms penetrates deeply into the opposite leaflet, giving rise to a tail-overlap region (Figure~\ref{fig:DMPC_interdigitations}). This provides a clear structural signature of interdigitation. We also observe pronounced differences among individual replicas under identical thermodynamic conditions, consistent with previously reported variability in mesoscopic properties such as area per lipid and membrane thickness (Figure~\ref{fig:DMPC_area_per_lipid_gauche_fraction}b and Figure~\ref{fig:DMPC_bilayerthickness}). These structural differences are readily apparent as well through visual inspection of the trajectories (Figure~\ref{fig:visu}).

\begin{figure}[htp]
    \centering
    \includegraphics[width=16
    cm]{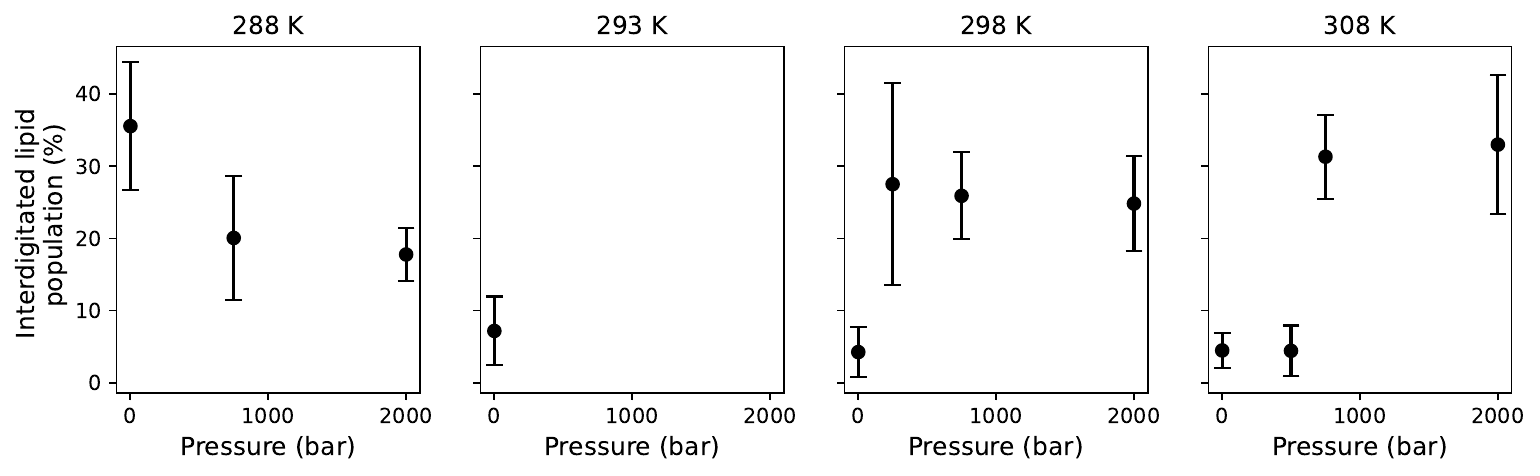}
    \caption{\textbf{Fraction of interdigitated lipids as a function of (P, T)}. The population of interdigitated lipids in DMPC bilayer at each temperature and pressure condition studied. Percentages presented here were calculated from the distributions of Figure~\ref{fig:DMPC_interdigitations_dist} as explained in Methods. Data points are the average among three replicas, with error bars corresponding to the SD among these replicas.}
    \label{fig:Interdigitations}
\end{figure}

To better quantify this phenomenon, we computed the distribution of distances between C14 atoms from the top leaflet and those from the bottom leaflet at matching $x$ and $y$ positions on a two-dimensional grid (Figure~\ref{fig:DMPC_interdigitations_dist}). In all cases, the distributions exhibit a peak around 0.3~nm. Under gel-phase conditions, they additionally display a long tail extending into negative values, corresponding to interdigitation, where atoms from the top leaflet extend below those from the bottom leaflet. To estimate the fraction of interdigitation, we fitted the positive-centered peak (non-interdigitated population) with a Gaussian function and report the results in Figure~\ref{fig:Interdigitations}, which call for the following observations.  

First, liquid phases are consistently associated with only a very small degree of interdigitation (typically below 10\%). In gel phases, however, we find that 20--40\% of the lipids are interdigitated. At 298~K and 308~K, there is no clear pressure dependence, whereas at 288~K the effect is less pronounced at 750~bar and 2000~bar, while the gel at 1~bar exhibits behavior comparable to that at higher temperatures and pressures.  

Second, the variability among replicas, reflected in the standard deviations, is substantial, especially for gel phases below 1000~bar. This indicates that while all replicas under a given (P,T) condition display similar values of the gauche fraction (Figure~\ref{fig:DMPC_area_per_lipid_gauche_fraction}a), the degree of interdigitation varies considerably. This variability can also be visualized in the distributions of membrane thickness across the $(x,y)$ plane (Figure~\ref{fig:Thickness_map}), which highlight the presence of long-lived interdigitation patterns that differ markedly among replicas. By contrast, in the liquid state the membrane thickness is uniform, and such patterns or variations among replicas are absent.

Notably, interdigitation appears to be a slow and only partially reversible process, potentially amplified by periodic boundary conditions, as the spatial extent (wavelength) of membrane undulations is constrained by the finite box size. Consequently, we consider this phenomenon to be partly an artifact of the simulation setup. As illustrated in Supplementary movies, interdigitation patterns of gel phases, which are reflected in the thickness distribution in the (x,y) place, vary among the replicas but are long-lived for a given replica, with little variations observed over a microsecond timescale. 

Experimentally, evidence for lipid interdigitation has been reported under specific conditions, such as high pressure or the presence of small headgroups and saturated chains\cite{yeagle_structure_2004}.
Interestingly, we note that the pressure/temperature dependence of interdigitation, which is reported as a color scale in Figure~\ref{fig:phase}, is consistent with experimental observations, with the presence of an intermediate region in the phase diagram corresponding to a ripple gel phase, more disordered and possibly interditigated.

\section{Conclusions}

In this work, we interrogate the pressure response of lipid bilayer that models the thickness of a bacterial membrane.
This question has both biophysical relevance, since pressure is an interesting thermodynamic variable for probing the conformational interplay between integral membrane proteins and the lipid bilayer, and biological significance, as such pressure conditions exist in natural environments on Earth. We compare two lipids: one, DMPC, that undergoes an experimentally observed phase transition in the investigated pressure--temperature $(P,T)$ range (288--308~K, 1--2000~bar), and another, $\Delta$9-cis-PC, that does not, owing to the presence of a single unsaturation per chain.

We first show that the CHARMM36 lipid force field yields results in excellent agreement with experimental data. In particular, the simulated liquid--gel transition line in the pressure--temperature diagram typically falls within 5--10~K and 100--300~bar of experimental measurements, which is remarkable given that these models were not explicitly parameterized to reproduce such thermodynamic properties. As expected, the control lipid $\Delta$9-cis-PC shows no phase transition in this range. We further compare two simulation protocols: an instantaneous switch in pressure/temperature versus a gradual ramp and find no significant differences in the resulting phases.  

Most of our simulations probe the liquid-to-gel transition. To explore possible hysteresis, we also study the reverse, melting transition. We observe mild hysteresis, especially when starting from a very stable gel at low temperature. It is however much more limited for points in the phase diagram that lie close to the transition line. 

We then analyze molecular markers of the phase transition across multiple length scales. Three descriptors (the area per lipid, membrane thickness, and acyl chain gauche fraction) are sensitive to phase changes. The gauche fraction emerges as an unambiguous reporter, since lipid tails adopt more trans conformations upon gelation and the gauche fraction distributions for the two phases are non-overlapping, with a threshold around 22\% gauche.
While area per lipid and membrane thickness are also widely used markers, they are strongly influenced by leaflet interdigitation (the interpenetration of the two bilayer leaflets). This leads to large variations between simulation replicas, depending on the degree of interdigitation, whereas the gauche fraction proves more robust.  

Within our simulation framework (finite size and timescale), we find that the interdigitated gel represents the thermodynamic equilibrium of the gel phase. However, the characteristic wavelength of this structure is constrained by the finite system size and periodic boundary conditions. As noted previously\cite{Khakbaz_2018}, the interdigitated phase is therefore not quantitatively equivalent to the ripple gel observed experimentally, which emerges at larger scales. We also note that interdigitated domains remain long-lived within the timescale of our simulations. We note a tendency towards less interdigitation at high pressures and low temperatures, which is reminiscent of the known transition between the ripple and lamellar gel phases, suggesting that the physical bases for this transition are also captured by the force field. A fully accurate description of that transition, however, would require larger simulation systems to accommodate the ripple phase, and extremely long time scales, or possibly dedicated enhanced sampling schemes.

This study demonstrates that all-atom molecular dynamics simulations not only reproduce experimental phase behavior but also provide mechanistic insights into how pressure and temperature affect model membranes. This reinforces the legitimacy of MD simulations as a tool to interpret and complement experimental measurements on these systems, as we have recently done\cite{Fourel2025}.
Building on this work, we are now combining experiments and simulations to examine the interplay between protein conformational dynamics and lipid organization in nanodiscs.
An exciting future direction is to investigate how these lipid environments, which pose unique challenges for simulations due to their large size and slow relaxation, differ from the infinite bilayers that are most commonly studied.

\section{Supporting Information Description}

\section{Acknowledgments}

This work was supported by the French National Research Agency under grant UnderPressure (ANR-22-CE29-0020).
YG and JH acknowledges support from the French National Research Agency under grant LABEX DYNAMO (ANR-11-LABX-0011).

\bibliography{main.bib}

%\section*{Table of Contents Image}

\clearpage

\section*{Supporting information}

PDF: Supplementary Figures.

\setcounter{figure}{0}
\renewcommand{\thefigure}{S\arabic{figure}}

\begin{figure}[htp]
    \centering
    \includegraphics[width=16cm]{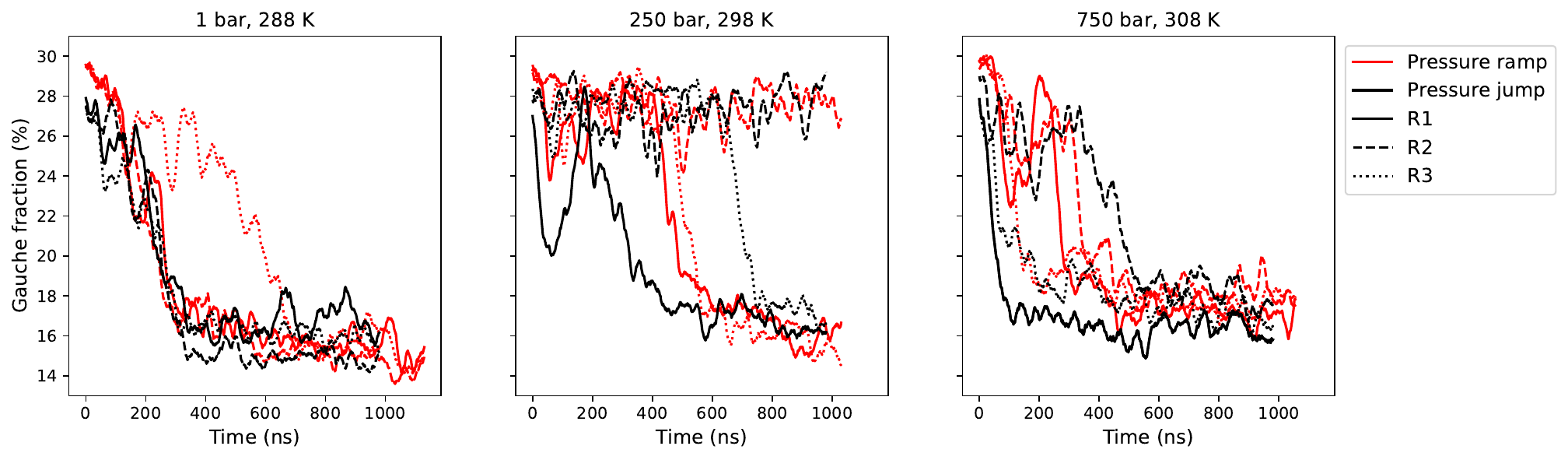}
    \caption{\textbf{Gauche fraction changes depending on the (P,T) schedule.} In our main simulations, (P,T) conditions were applied as instantaneous jump (black curves). We tested the sensitivity of the transition to the (P,T) schedule by applying linear ramps in T and P (red curves). Three conditions were tested : 1~bar and 288~K (left panel), 250~bar and 298~K (middle panel), and 750~bar and 308~K (right panel). Each line style represents to a replica. Gauche fraction values are shown as running averages over 20~ns.}
    \label{fig:PT_ramp}
\end{figure}

\vspace{2cm}

\begin{figure}[htp]
    \centering
    \includegraphics[width=16cm]{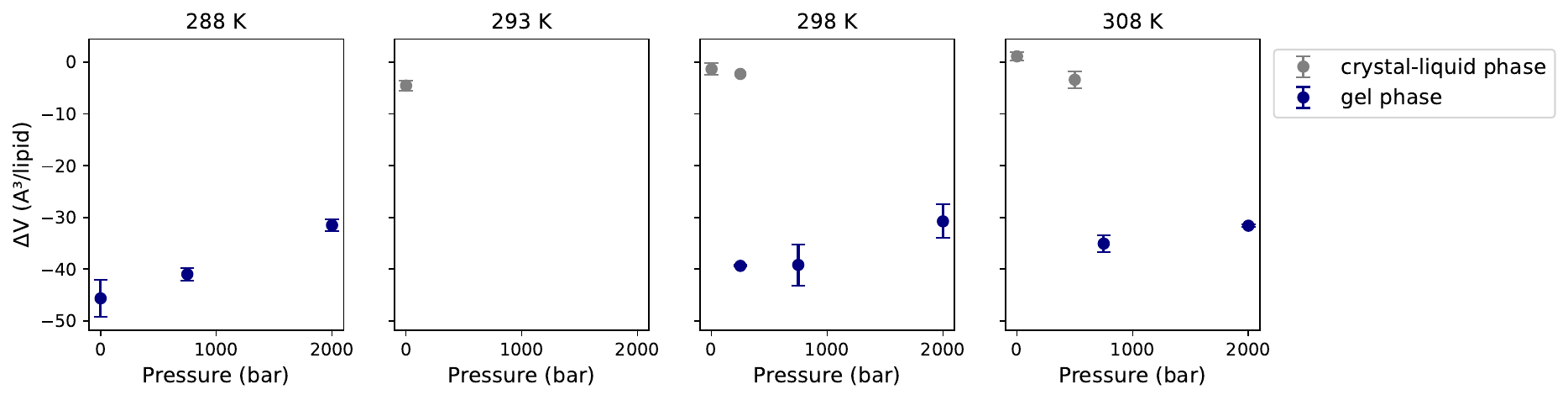}
    \caption{\textbf{Fluid-to-gel volume change $\Delta$V.} The variation in the volume of the simulation box is shown between the mean volume calculated during first 5~ns of simulation (without considering the volume drop due to the instantaneous pressure jump) and the mean volume over the last 300~ns of the simulation. The data points correspond to the average among three replicas, and the error bars correspond to the SD among these replicas. Note that at 250~bar at T=298~K, two replicas underwent gelation and one replica remained fluid; they are represented separately.}
    \label{fig:DV_perlipid}
\end{figure}

\begin{figure}[htp]
    \centering
    \includegraphics[width=14cm]{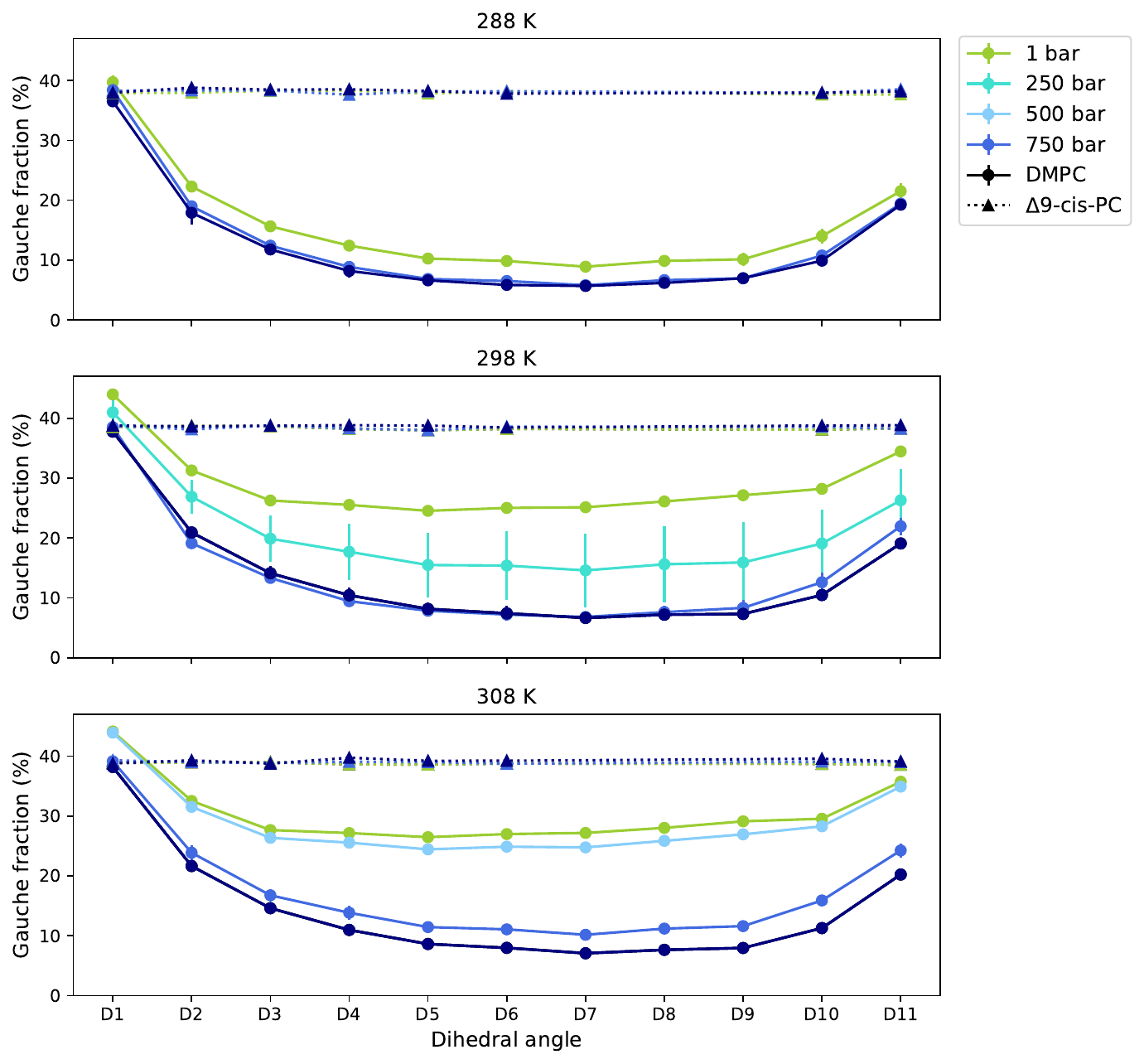}
    \caption{\textbf{Gauche fraction along the hydrophobic tails (DMPC, $\Delta$9-cis-PC)}.  The gauche fraction of each dihedral of DMPC and $\Delta$9-cis-PC hydrophobic tails is presented here, calculated from the last 300~ns of the simulation. All simulation T,P conditions are presented. Data points are the average among three replicas for DMPC (a single replica was simulated for $\Delta$9--cis--PC), with error~bars corresponding to the SD among these replicas.}
    \label{fig:Gauche_fraction_by_dihedral}
\end{figure}

\begin{figure}[htp]
    \centering
\includegraphics[width=14cm]{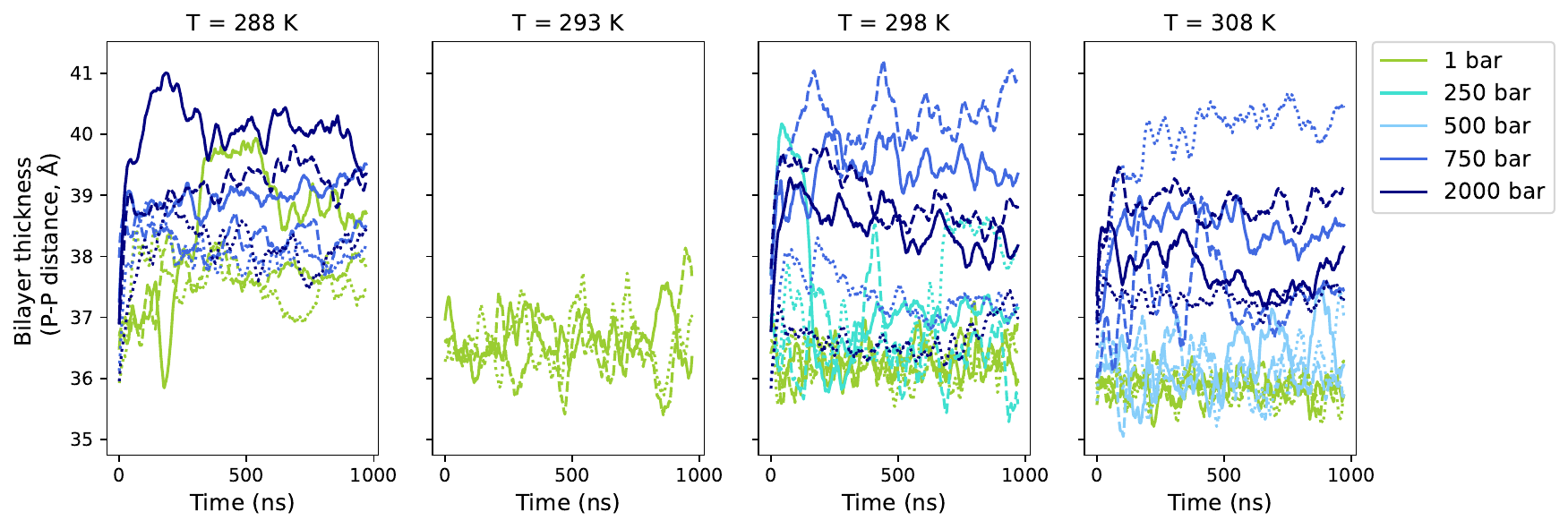}
\caption{\textbf{DMPC bilayer thickness timeline} The DMPC bilayer thickness as a function of time is shown. The bilayer thickness was calculated as the distance between the phosphorus atoms from the top leaflet and the phosphorus atoms from the bottom leaflet. Each panel corresponds to a different simulated temperature (288~K, 293~K, 298~K and 308~K). The simulated pressures at each temperature are shown on the corresponding panels. The values for each replicas are represented by a different line style with replica 1 in solid, replica 2 in dashed, and replica 3 in dotted style. Values presented here correspond to the running averages computed over 30~ns.}
\label{fig:DMPC_bilayerthickness}
\end{figure}

\begin{figure}[htp]
    \centering
    \includegraphics[width=14cm]{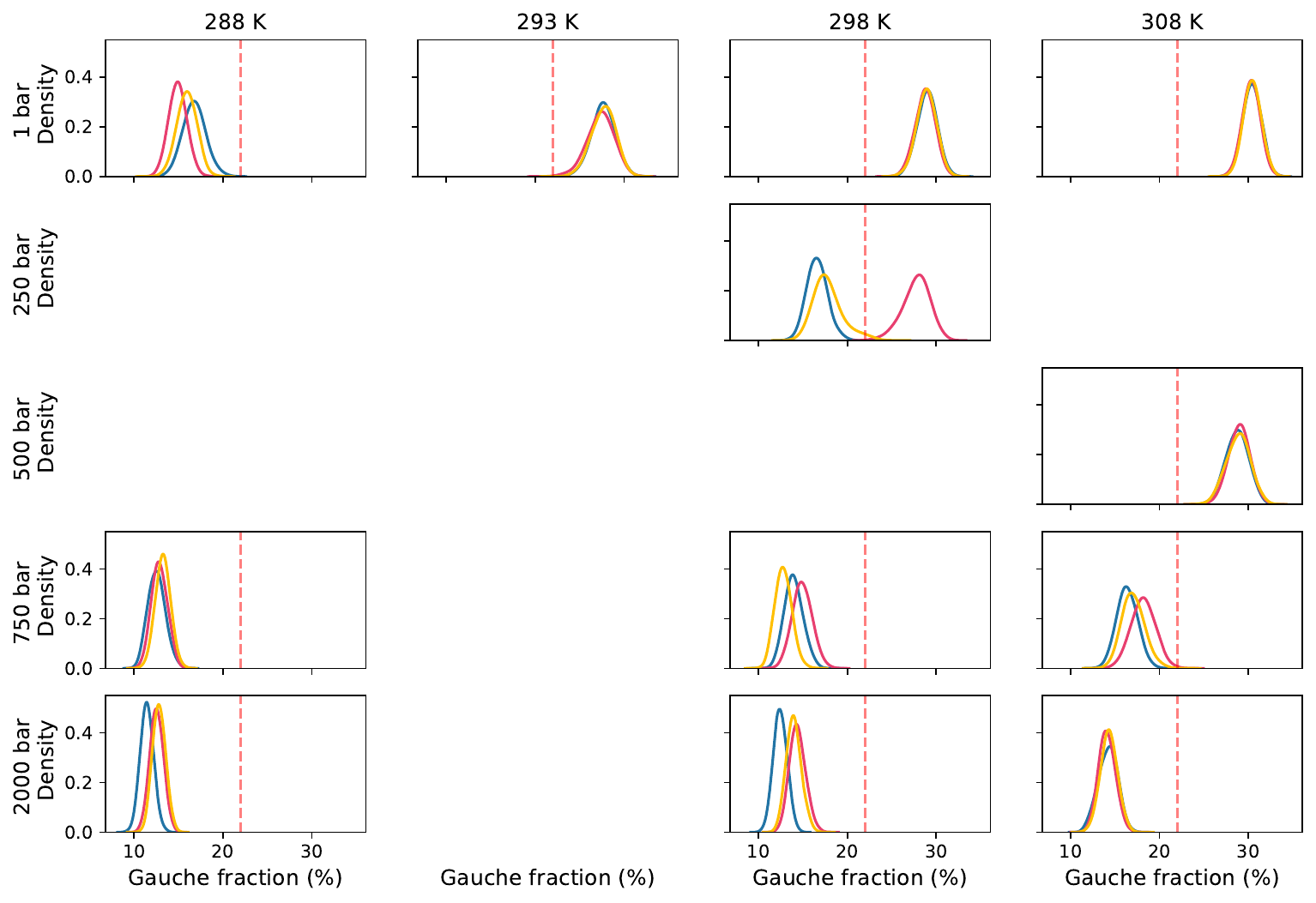}
    \caption{\textbf{Distributions of gauche fraction values}. The distribution of the gauche fraction values calculated over the entire lipid population over the over the last 300~ns of simulation are shown here at every T,P condition. The distribution for each replica can be distinguished by the colour of the line with replica 1 in blue, replica 2 in magenta, and replica 1 in gold. The red dashed line indicates the boundary between the gauche fraction values of crystal-liquid phase and the gel phase.}
    \label{fig:DMPC_gauche_frac_distrib}
\end{figure}

\begin{figure}[htp]
    \centering
    \includegraphics[width=14cm]{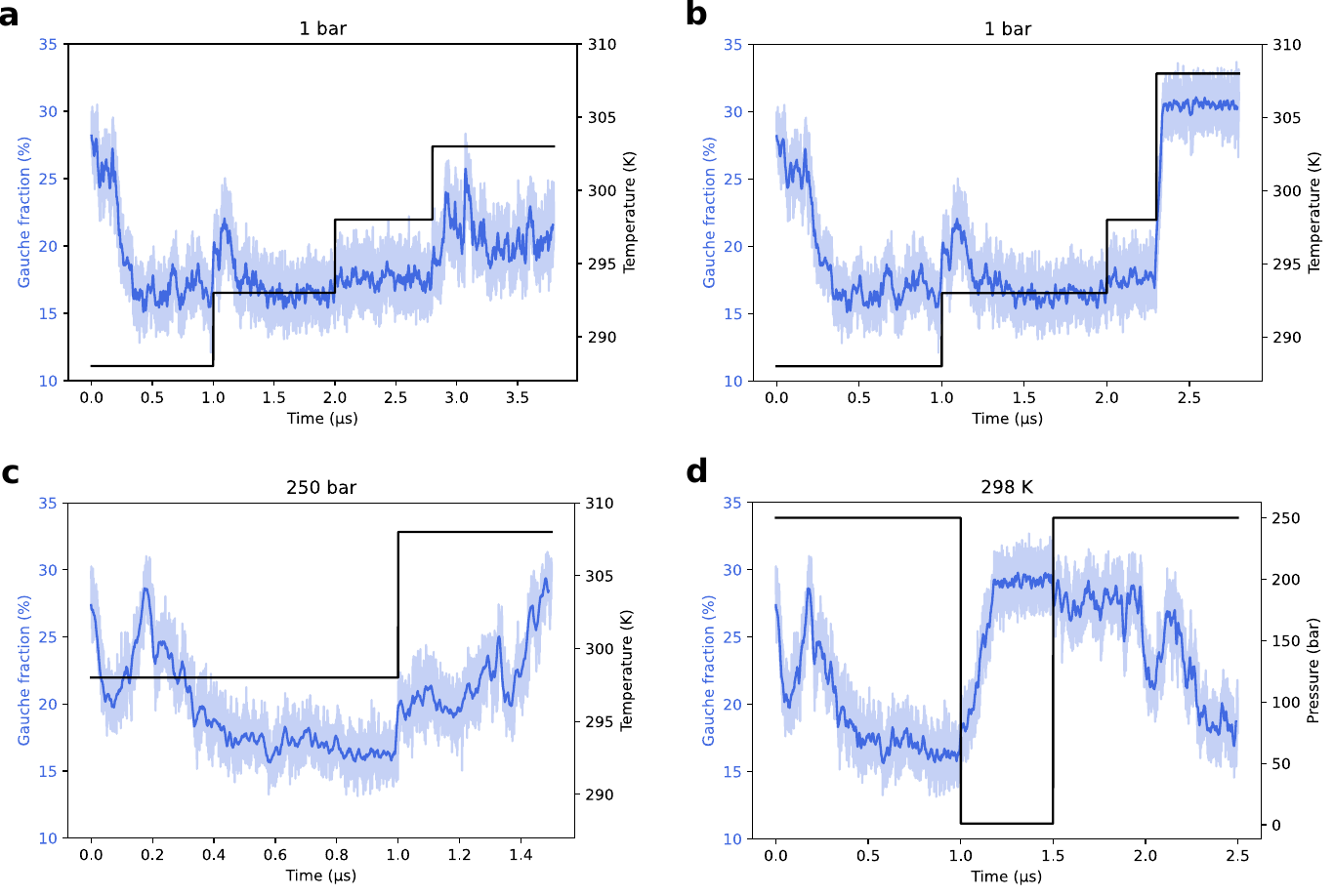}
    \caption{\textbf{Reversibility of the fluid-to-gel transition.}
    The reverse, gel-to-fluid transition was investigated by increasing the temperature (a),(b) and (c), or decreasing the pressure (d). The gauche fraction was used as an indicator of the lipid phase, with a high gauche fraction value corresponding (above 22\%, Figure~\ref{fig:DMPC_gauche_frac_distrib}) to the fluid phase and a low gauche fraction value (i.e., a value below 22\%, Figure~\ref{fig:DMPC_gauche_frac_distrib}) corresponding to the gel phase. (a),(b) To transition from a gel to a fluid by increasing the temperature, we started at 1~bar and 298~K, then increased the temperature to 303~K, applying temperature jumps that passed through 293~K for 1 µs, 298~K for 800~ns and 303~K for 1~µs. We need to heat the bilayer up to 308~K (b) to obtain the crystal-liquid phase. (c) At 250~bar 298~K the DMPC bilayer goes from fluid to gel. By heating the bilayer up to 308~K we go back to the crystal liquid phase. (c) At 250~bar and 298~K, the DMPC bilayer undergoes a fluid-to-gel phase transition. To probe the hysteresis of this transition the pressure was decreased to 1~bar. To verify the reversibility of the transition we applied back a pressure of 250~K and saw that the lipids are transiting from fluid-to-gel again. The gauche fraction of the entire lipid population was calculated, with the navy blue line representing the smoothed gauche fraction over 10~ns. This study was done with one of the replicas.}
    \label{fig:hysteresis}
\end{figure}

\begin{figure}[htp]
    \centering
    \includegraphics[width=10cm]{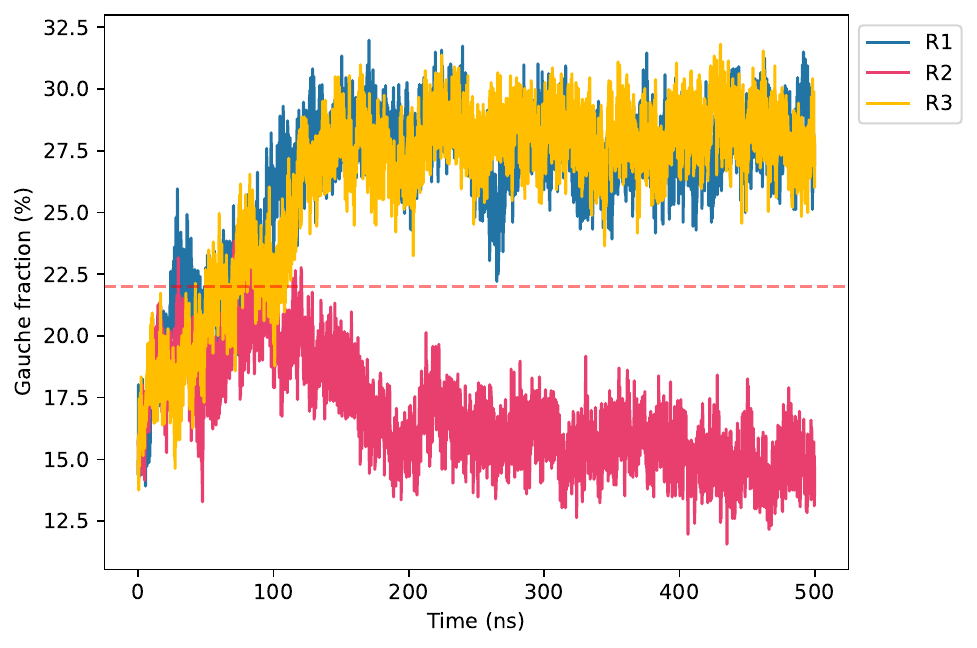}
    \caption{\textbf{Split-state starting point at 1~bar and 293~K to test for phase coexistence.} 
    Three replica simulations were started from a hybrid gel/fluid bilayer: a disc of gel-phase DMPC surrounded by the liquid-crystal phase.
    The gauche fraction as a function of time at 1~bar and 293~K is shown for each replica. The red dashed line indicates the boundary between the gauche fraction values corresponding to the crystal-liquid (above) and gel phase (below), see Figure \ref{fig:DMPC_gauche_frac_distrib}}.
    \label{fig:DMPC_phase_293K}
\end{figure}

\begin{figure}[htp]
    \centering
\includegraphics[width=14cm]{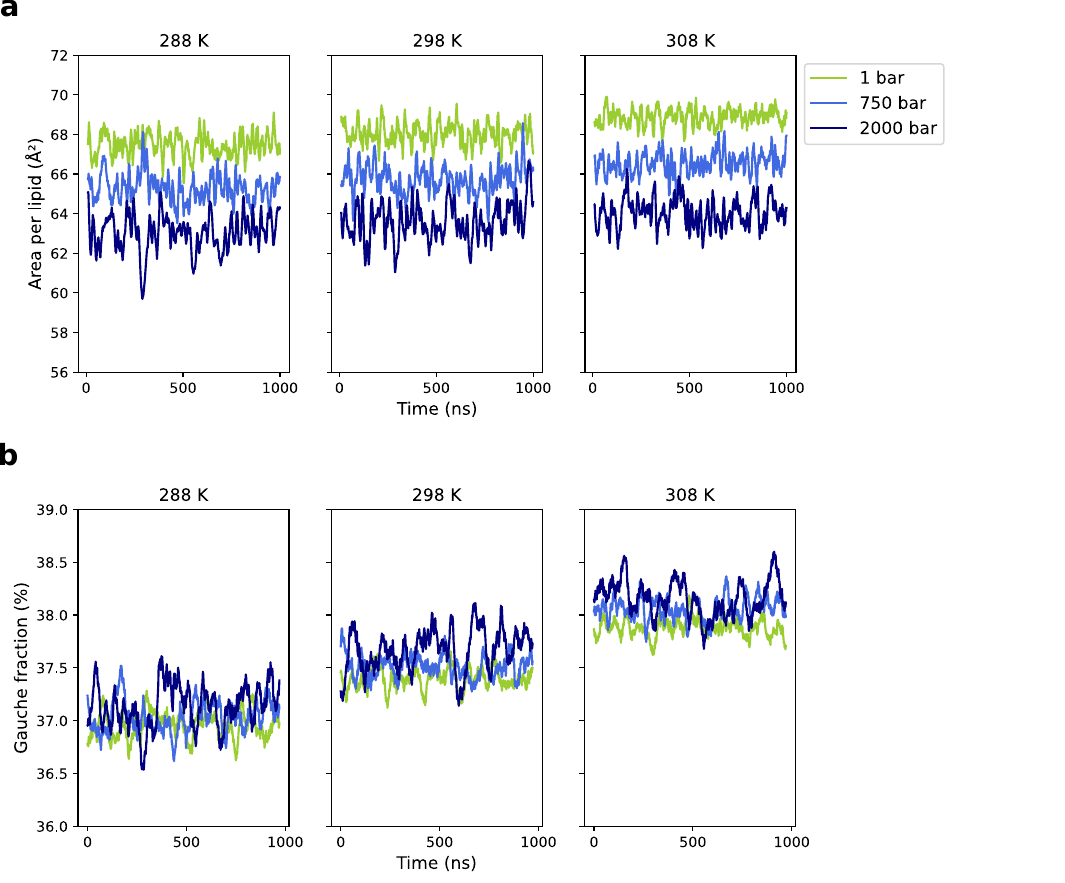}
\caption{\textbf{Dynamics of the (P,T) response of the negative control, $\Delta$9-cis-PC}. (a) The area per lipid as function of time for each T,P condition simuated.(b) Gauche fraction of the dihedral angles of the tails of the $\Delta$9-cis-PC lipids as a function of time. For both figure, each panel corresponds to a simulated temperature (288~K, 298~K and 308~K) with the corresponding different simulated pressures shown (1~bar, 750~bar and 2000~bar). Area per lipid : smoothing window : 10~ns; gauche fraction : 30~ns}
\label{fig:D9_area_per_lipid_gauche_fraction}
\end{figure}

\begin{figure}[htp]
    \centering
    \includegraphics[width=14cm]{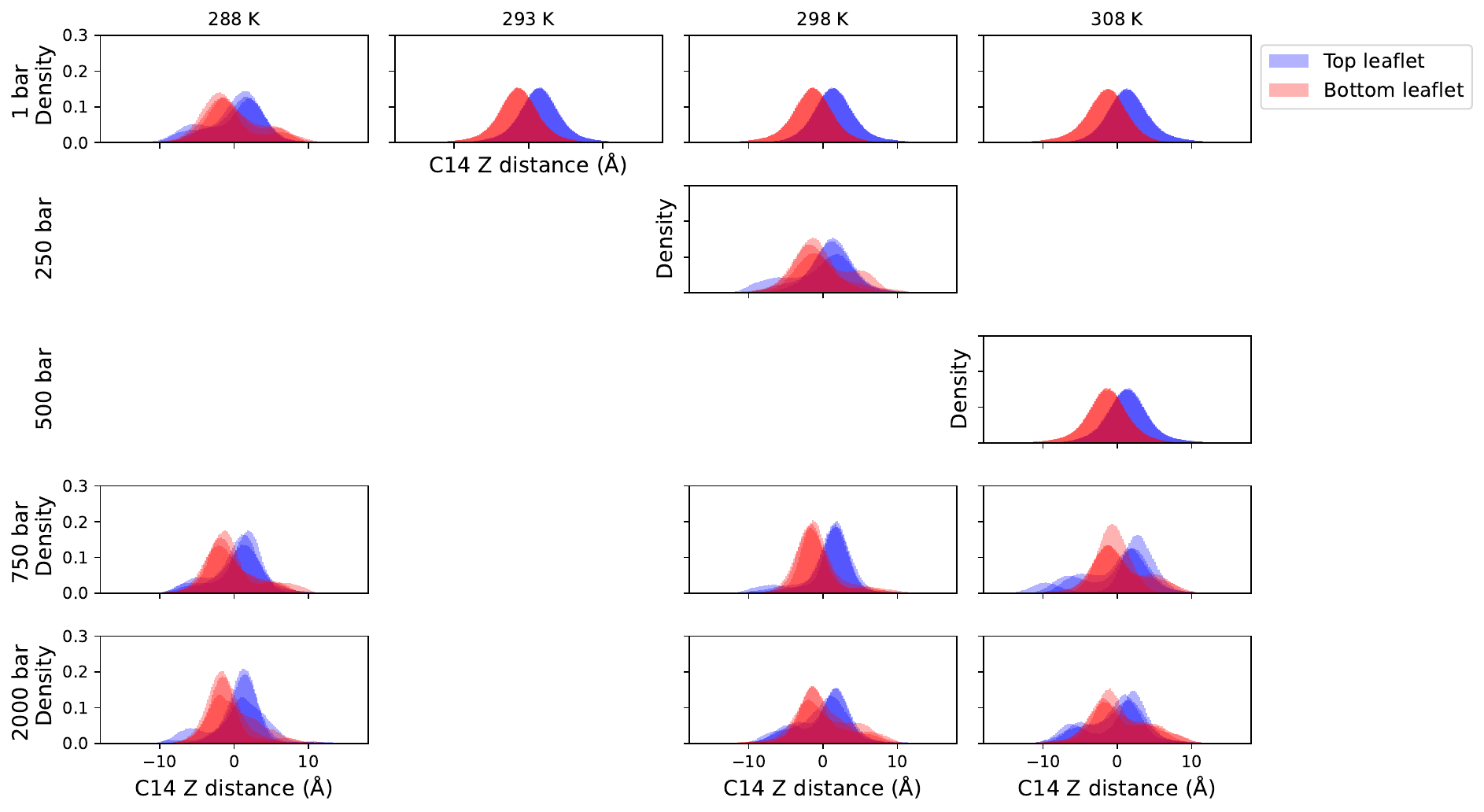}
    \caption{\textbf{Distribution of C14 z-positions (DMPC)}. Distribution of the z-position values of the last carbon of each DMPC tail (C14) with the distinction between lipids of the top leaflet (in blue) and those in the bottom leaflet (in red). Values presented here were taken from the last 300~ns of the simulation. All simulated T,P conditions are presented. The different replicas are represented by a shading of the color.}
    \label{fig:DMPC_interdigitations}
\end{figure}

\begin{figure}[htp]
    \centering
    \includegraphics[width=14
    cm]{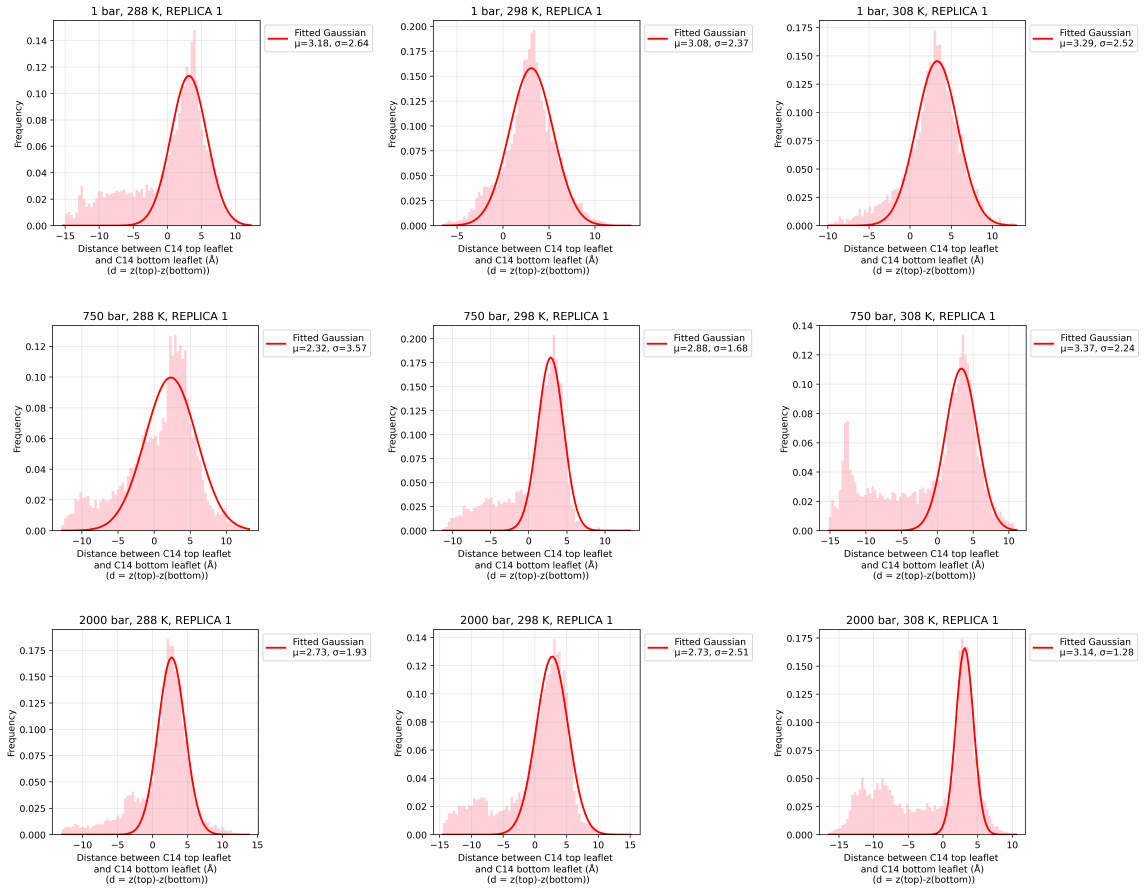}
    \caption{\textbf{A metric to quantify bilayer interdigitation}. The distribution of the distance between the last carbon the each DMPC tail (C14) of the top and bottom leaflets, interpolated on a grid, is shown in pink (see Methods). A Gaussian distribution was fitted to the distribution of distances, represented here by the red curve. The fraction of the distribution that us not accounted for by the Gaussian represents the interdigitated fraction of the bilayer. Values presented here were taken from the last 300~ns of the simulation. All simulation T,P conditions are presented. Only 1 replica is represented here (Replica 1).}
    \label{fig:DMPC_interdigitations_dist}
\end{figure}

\begin{figure}[htp]
    \centering
    \includegraphics[width=14cm]{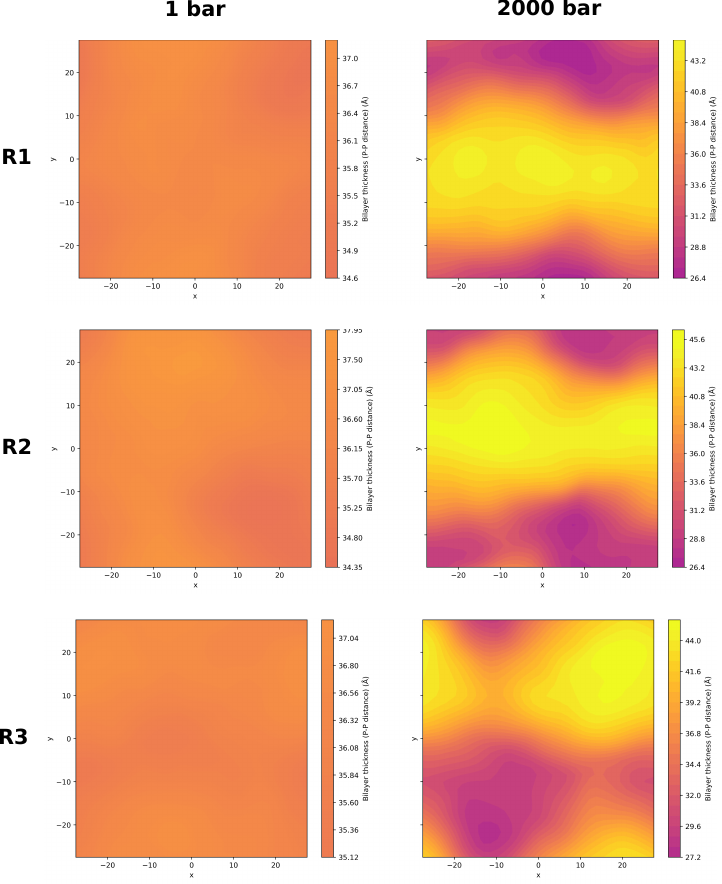}
    \caption{\textbf{DMPC Bilayer thickness maps}. The bilayer thickness is shown here with the detail of each replica, 1 (first row), 2 (middle row) and 3 (last row). The bilayer thickness was calculated as the distance between the lipid's head P atoms from the top and bottom leaflet, interpolated on a grid. These plots represent data aggregated over the last 300~ns of the simulation. Simulated pressures are shown on the corresponding panels (for each condition we compare 1~bar and 2000~bar at 298~K).}
    \label{fig:Thickness_map}
\end{figure}

\end{document}